\documentclass[letterpaper, 10 pt, conference]{ieeeconf}  

\IEEEoverridecommandlockouts                              

\usepackage{cite}
\usepackage{amsmath,amssymb,amsfonts}
\usepackage{algorithmic}
\usepackage{graphicx}
\usepackage{textcomp}

\newtheorem{proposition}{Proposition}

\newtheorem{problem}{Problem}
\usepackage{xcolor}
\usepackage{cite}
\usepackage[colorlinks=true,allcolors=steelblue]{hyperref}
\definecolor{steelblue}{RGB}{70,130,180}
\def\bbn{\mathbb N}
\def\bbz{\mathbb Z}
\def\bbr{\mathbb R}

\DeclareMathOperator*{\argmin}{argmin}

\newcommand{\abs}[1]{\left |#1 \right\vert}

\title{\LARGE \bf Distributed Implementation of Minimax Adaptive Controller \\ For Finite Set of Linear Systems}

\author{Venkatraman Renganathan, Anders Rantzer, and Olle Kjellqvist
\thanks{This project has received funding from the European Research Council (ERC) under the European Union’s Horizon 2020 research and innovation program under grant agreement No 834142 (Scalable Control). The authors are with the Department of Automatic Control LTH, Lund University, Sweden. (e-mail: (venkatraman.renganathan,anders.rantzer, olle.kjellqvist)@control.lth.se).}
}

\begin{document}

\maketitle
\thispagestyle{empty}
\pagestyle{empty}

\begin{abstract}

This paper deals with a distributed implementation of minimax adaptive control algorithm for networked dynamical systems modeled by a finite set of linear models. To hedge against the uncertainty arising out of finite number of possible dynamics in each node in the network, it collects only the historical data of its neighboring nodes to decide its control action along its edges. This makes our proposed distributed approach scalable. Numerical simulations demonstrate that once each node has sufficiently estimated the uncertain parameters, the distributed minimax adaptive controller behaves like the optimal distributed H-infinity controller in hindsight.

\end{abstract}

\section{Introduction}
\label{sec:introduction}
Control of large-scale and complex systems is often performed in a distributed manner \cite{antonelli2013interconnected} as it is practically difficult for every agent in the network to have access to the global information about the overall networked system while deciding its control actions. On the other hand, designing optimal distributed control laws when the networked system dynamics are uncertain still remains an open problem. This naturally calls for a learning-based controller to be employed in such uncertain settings and it was shown in \cite{astromwittenmark} that an adaptive controller can learn the true system dynamics online through sufficient parameter estimation and then control the system. Multiple model-based adaptive control problems have been extensively studied in the control literature \cite{anderson2000multiple, hespanha2001multiple, anderson2001multiple, kuipers2010multiple, narendra1997adaptive, narendra2011changing}. The minimax problem of handling adversarial disturbance and the uncertain parameters leading to an unknown linear system was introduced in \cite{didinsky1994minimax}. It was specialized to scalar systems with unknown sign for input matrix in \cite{rantzer2020minimax}, to finite sets of linear systems in \cite{rantzer2021minimax}. However, designing optimal distributed control laws that address the uncertainty prevailing over true model of the networked system out of the finite set of linear models still remains an open problem.  

Minimax adaptive control problems are challenging mainly due to the exploration and exploitation trade-off that inevitably comes with the learning and controlling procedure. Aiming for a distributed implementation on top of it complicates things further. However, there are certain classes of systems for which scalable implementation of distributed minimax adaptive control is possible, such as the the one with linear time-invariant discrete time systems with symmetric and Schur state matrix. Such system models are common in irrigation networks. For instance, the authors in \cite{vladu2022decentralized} 
computed a closed-form expression for a decentralised $H_{\infty}$ optimal controller with diagonal gain matrix for network systems with acyclic graphs. Similarly, the authors in \cite{lidstrom2017h} computed a closed-form expression for the distributed $H_{\infty}$ optimal state feedback law for systems with symmetric and Schur state matrix, where the total networked system comprised of subsystems with local dynamics, that share only control inputs and each control input affecting only two subsystems.  

\textit{Contributions:} We extend the problem setting in \cite{lidstrom2017h} by considering finite number of possible local dynamics in each node and control action along each edge. Our main contributions in this paper are as follows:
\begin{enumerate}
    \item We develop scalable and distributed implementation of minimax adaptive control for networked systems where each node in the network is required to maintain the history of just its own neighboring nodes to hedge against the uncertainty in its local system dynamics.
    \item We demonstrate the effectiveness of our proposed approach using a buffer network based numerical example to advocate the use of scalable and distributed minimax adaptive control law for finite set of linear systems.
\end{enumerate}
Following a short summary of notations, this paper is organized as follows: In \S\ref{sec:problemformulation}, the main problem formulation of distributed minimax adaptive controller and its distributed implementation is presented. The effectiveness of the proposed algorithm is then demonstrated in \S\ref{sec:simulation}. Finally, the paper is closed in \S\ref{sec:conclusion} along with a summary of results and directions for future research.

\section*{Notations}
The cardinality of the set $A$ is denoted by $\left | A \right \vert$. The set of real numbers, integers and the natural numbers are denoted by $\bbr, \bbz, \bbn$ respectively. The subset of real numbers greater than a given constant say $a \in \bbr$ and between two constants $a, b \in \bbn$ with $a < b$ are denoted by $\bbr_{> a}$ and $[a:b]$ respectively. A vector of size $n$ with all values being one is denoted by $\mathbf{1}_{n}$. For a matrix $A \in \bbr^{n \times n}$, we denote its transpose and its trace by $A^{\top}$ and $\mathbf{Tr}(A)$ respectively. We denote by $\mathbb{S}^{n}$ the set of symmetric matrices in $\bbr^{n \times n}$ and the set of positive (semi)definite matrices as $\mathbb{S}^{n}_{++} (\mathbb{S}^{n}_{+})$.
A symmetric matrix $P \in \mathbb{S}^{n}$ is said to be positive definite (positive semi-definite) if for every vector $x \in \bbr^{n} \backslash \{0\}$, we have $x^{\top} P x > 0 \, \, (x^{\top} P x \geq 0)$ and is denoted by $P \succ 0 (P \succeq 0)$ . An identity matrix in dimension $n$ is denoted by $I_{n}$. Given $x \in \bbr^{n}, A \in \bbr^{n \times n}, B \in \bbr^{n \times n}$, the notations $\abs{x}^{2}_{A}$ and ${\left \| B \right \Vert}^{2}_{A}$ mean $x^{\top} A x$ and $\mathbf{Tr}\left(B^{\top} A B \right)$ respectively. We model the networked dynamical system in discrete time setting with a graph $\mathcal{G}$ comprising a node set $\mathcal{V}$ representing $|\mathcal{V}| = N$ subsystems and edge set $\mathcal{E} \subset \mathcal{V} \times \mathcal{V}$ representing a set of $|\mathcal{E}| = E$ communication links amongst the subsystems. The incidence matrix encoding the edge set information is denoted by $\mathcal{I} \in \mathbb{R}^{N \times E}$. We denote by $\mathcal{N}_{i} = \{j \in \mathcal{V} : (j, i) \in \mathcal{E}\}$ the neighbor set of agent $i$ at time $t$, whose states are available to agent $i$ via communication links. The degree of $i$ is denoted as $d_{i} := \abs{\mathcal{N}_{i}}$.


\section{Problem Formulation}
\label{sec:problemformulation}
\subsection{Distributed Minimax Adaptive Control Problem}
We associate with each node $i \in \mathcal{V}$, a state $x_{i}(t) \in \mathbb{R}$ at time $t \in \mathbb{Z}_{\geq 0}$. Each subsystem $i \in \mathcal{V}$ updates its own states through the interaction with its neighbors as 
\begin{align} \label{eqn_node_dynamics}
    x_{i}(t+1) = a_{i} x_{i}(t) + b \sum_{j \in \mathcal{N}_{i}} (u_{i}(t) - u_{j}(t))  + w_{i}(t),
\end{align}
where $a_{i} \in (0,1), b > 0$. A control policy is termed as \emph{admissible} if it is stabilising and has causal implementation. The control input of node $i \in \mathcal{V}$ is 
\begin{align} \label{eqn_control_policy}
    u_{i}(t) = \pi_{t}\left( x(0), x(1), \dots, x(t), u(0), \dots, u(t-1) \right),
\end{align} 
with the control policy $\pi_{t} \in \Pi$ and $\Pi$ denoting the set of all admissible control policies. The control input between two nodes $(i,j) \in \mathcal{E}$ is such that what is drawn from subsystem $j$ is added to subsystem $i$. The additive disturbance $w_{i}(t) \in \mathbb{R}$ affects the node $i \in \mathcal{V}$. Note here that the dynamics of each node in the network is coupled with the other nodes only through their control inputs and has to satisfy the communication constraints 
\begin{align} \label{eqn_communication_condition}
a^{2}_{i} + 2 b^{2} d_{i} < a_{i}, \quad \forall i \in \mathcal{V}.
\end{align} 
In the following problem setting considered in this paper, the true system model $a_{i}$ governing the dynamics of node $i \in \mathcal{V}$ in \eqref{eqn_node_dynamics} is \emph{unknown}. However, we have the following two assumptions for the considered problem setting. 
\begin{enumerate}
    \item The dynamics in \eqref{eqn_node_dynamics} defined by $a_{i}$ is assumed to belong to a finite set $\mathbf{A_{i}}$ with $\abs{\mathbf{A_{i}}} = M \in \mathbb{N}_{\geq 1}$. That is, 
    \begin{equation} \label{eqn_node_uncertainty}
        a_{i} \in \mathbf{A_{i}} := \{ a_{i} \in (0,1) \mid \eqref{eqn_communication_condition} \text{ is satisfied} \}.
    \end{equation}
    \item The dynamics of each node $i \in \mathcal{V}$ is independent of the dynamics of its neighbor $j \in \mathcal{N}_{i}$. This means that the choice of any one of the $M$ values of $a_i \in \mathbf{A_{i}}$ being the true $a_{i}$ does not affect the choice of any one of the $M$ values of $a_j \in \mathbf{A_{j}}$ being the true $a_{j}$ for every neighbor $j \in \mathcal{N}_{i}$. 
\end{enumerate}
We can concatenate the states of all nodes in \eqref{eqn_node_dynamics} as $x(t) = \begin{bmatrix} x^{\top}_{1}(t) & x^{\top}_{2}(t) & \dots & x^{\top}_{N}(t) \end{bmatrix}^{\top} \in \mathbb{R}^{N}$. Similarly, $u(t) = \{u_{i}(t)  - u_{j}(t)\}_{(i,j) \in \mathcal{E}} \in \mathbb{R}^{E}$ denotes the concatenated control inputs.

\begin{problem} \label{problem_1}
Let $\gamma > 0$. Given the uncertainty set $\mathbf{A_{i}}$ for every node $i \in \mathcal{V}$, find a control policy $\pi_{t} \in \Pi, \forall t$ such that the control input at time $t$ for every node $i \in \mathcal{V}$ given by \eqref{eqn_control_policy} satisfies the inequality
\begin{align}
\label{eqn_cost}
\sum^{T}_{\tau=0} \sum_{i \in \mathcal{V}} \left( \abs{x_{i}(\tau)}^{2} + \abs{u_{i}(\tau)}^{2} \right) \leq \gamma^{2} \sum^{T}_{\tau=0} \sum_{i \in \mathcal{V}} \abs{w_{i}(\tau)}^{2},
\end{align}
for all solutions of  \eqref{eqn_node_dynamics} with uncertainty in its dynamics given by \eqref{eqn_node_uncertainty} and communication constraints defined using \eqref{eqn_communication_condition} and all time horizon $T \geq 0$.
\end{problem}


\subsection{Distributed Implementation}
\label{sec:analysis}
This section describes the distributed implementation of the minimax adaptive control algorithm given the set $\mathbf{A_{i}}$ for every node $i \in \mathcal{V}$ is known. There exists an one-to-one correspondence between Problem \ref{problem_1} and the centralised minimax adaptive control problem setting given in \cite{rantzer2021minimax}. However, the number of possible system matrices $A \in \mathbb{R}^{N \times N}$ formed by concatenating the dynamics of all nodes in \eqref{eqn_node_dynamics} grows exponentially in terms of the network size as $N^{M}$ due to the uncertainty in dynamics of each node. To hedge against the uncertainty in the system matrix $A$, it is natural for any controller to consider collecting historical data. Given that each node $i \in \mathcal{V}$ in the network has access to only local information from its neighbors, the process of hedging against the uncertainty prevailing over its dynamics $a_{i} \in \mathbf{A_{i}}$ involves collecting historical data from its local neighbors $\mathcal{N}_{i}$ to arrive at its own control action. The following proposition provides a distributed implementation of the controller that solves Problem \ref{problem_1}.  

\begin{proposition} \label{proposition_1}
If Problem \ref{problem_1} is solvable, then it has a solution for every node $i \in \mathcal{V}$ of the form
\begin{align} 
    u_{i}(t) &= \frac{b x_{i}(t)}{a^{\dagger}_{i}(t) - 1}, \quad \text{where}, \label{eqn_node_minimax_controls}
\end{align}
\begin{equation}
\footnotesize
    a^{\dagger}_{i}(t) = \argmin_{a_{i} \in \mathbf{A_{i}}} \sum^{t-1}_{\tau=0} \abs{a_{i} x_{i}(\tau) + b \sum_{j \in \mathcal{N}_{i}} (u_{i}(\tau) - u_{j}(\tau)) - x_{i}(\tau+1)}^{2}.
\end{equation}
\end{proposition}
\noindent \textbf{Remarks:} Every node $i \in \mathcal{V}$ selects the model that best fits the disturbance trajectory modelled using the collected history in a least-square sense. Note that the Proposition \ref{proposition_1} has only been verified for values of $\gamma$ that satisfies the Riccati type inequality given by Theorem 3 in \cite{rantzer2021minimax}. Such a detailed exposition about the least achievable gain bound is actively being carried out in our ongoing future work and this paper is just concerned about the distributed implementation. Note that for the proposed implementation of the distributed minimax adaptive controller, the knowledge of the gain bound is not needed, meaning that we need not search for the best value of $\gamma$ in the space of exponential number of corner matrices.


\section{Numerical Simulation}
\label{sec:simulation}

\subsection{Simulation Setup}
\begin{figure}
\centering
\includegraphics[scale=0.085]{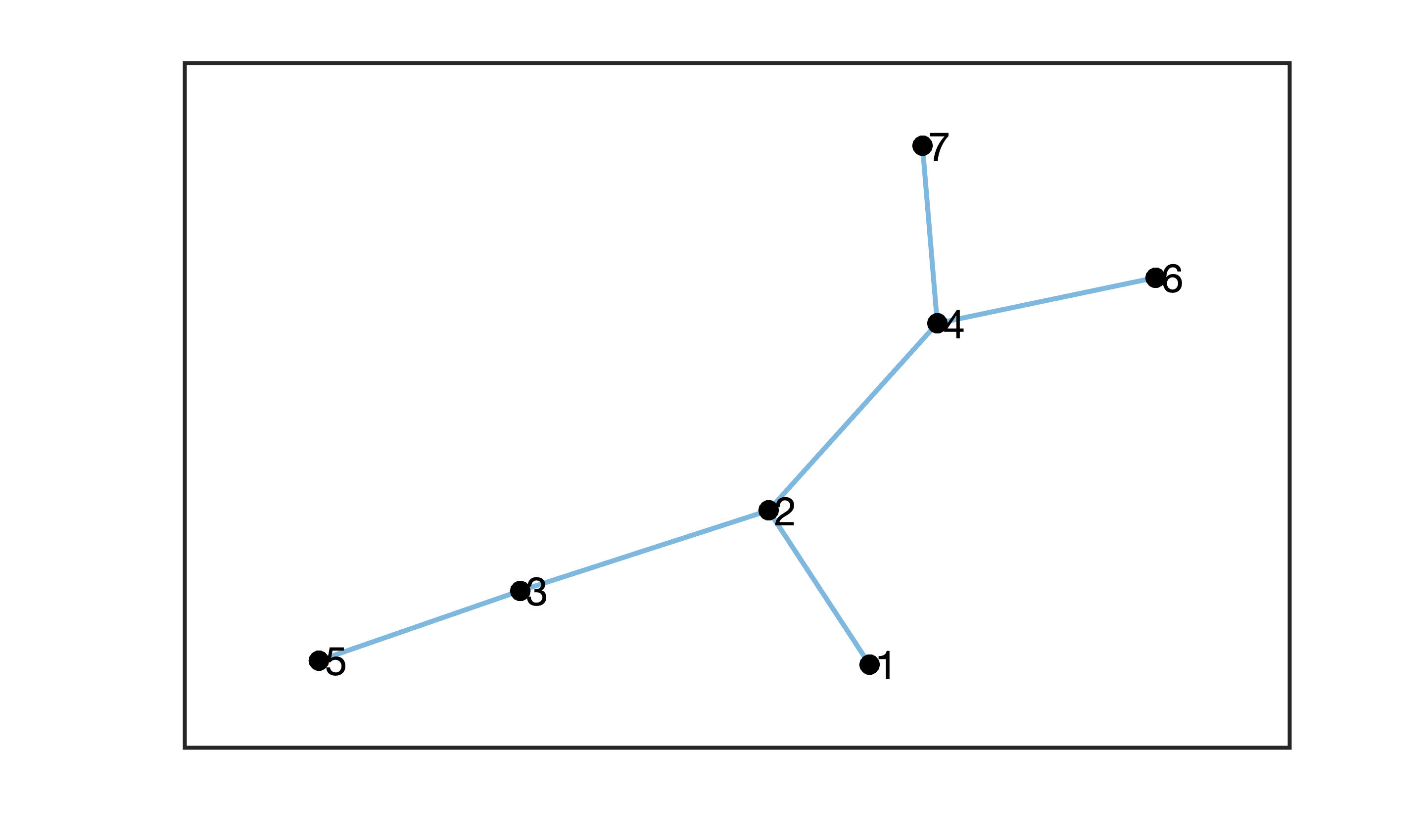}
\caption{The underlying graph of the networked system considered in the numerical simulation with $V = 7$ nodes is shown here.}
\label{fig_network}
\end{figure}

To demonstrate our proposed approach, we consider a buffer network with $N = 7$ nodes as shown in the figure \ref{fig_network}. We let $b = 0.1$ so that the input matrix for the whole network is simply the scaled incidence matrix, $B = bI$, with the incidence matrix $\mathcal{I}$ being
\begin{align*}
\mathcal{I} = \begin{bmatrix}
        -1 & 0 & 0 & 0 & 0 & 0 \\
        1 & -1 & -1 & 0 & 0 & 0 \\
        0 & 1 & 0 & -1 & 0 & 0 \\
        0 & 0 & 1 & 0 & -1 & -1 \\
        0 & 0 & 0 & 1 & 0 & 0 \\
        0 & 0 & 0 & 0 & 1 & 0 \\ 
        0 & 0 & 0 & 0 & 0 & 1  
    \end{bmatrix}.
\end{align*}
Five different models $(M = 5)$ were generated randomly for each node $i \in \mathcal{V}$ satisfying \eqref{eqn_communication_condition}. The total time horizon was set to be $T = 30$. To simulate the above system with a disturbance signal that has infinite $L_{2}$ norm (energy), we chose the sinusoidal disturbance. One of the $M$ possible models in the set $\mathbf{A_{i}}$ was randomly picked and fixed to be the true system model governing its dynamics for every node $i \in \mathcal{V}$  throughout the time horizon. For every node $i \in \mathcal{V}$, the distributed minimax adaptive control inputs were computed using \eqref{eqn_node_minimax_controls} and the distributed $H_{\infty}$ control inputs were computed as described in \cite{lidstrom2017h} using the true $a_{i} \in \mathbf{A_{i}}$ as
\begin{align} \label{eqn_Hinfty_controls}
    u^{\star}_{i}(t) &= \frac{bx_{i}(t)}{a_{i} - 1}. 
\end{align}
In the last simulation, we compute a nominal distributed $H_{\infty}$ controller where each node $i \in \mathcal{V}$ computes its control input given by \eqref{eqn_Hinfty_controls} by assuming an $a_{i}$ value satisfying \eqref{eqn_communication_condition} to be its constant nominal value $\forall t \in \mathbb{N}$. Note that any value in $(0,1)$ can be termed as the nominal value for $a_{i}$. We compare the system trajectory obtained using the nominal distributed $H_{\infty}$ controller for the true system against the system trajectory from the distributed $H_{\infty}$ controller which knows the true $a_{i}$ for all nodes $i \in \mathcal{V}$. This is to qualitatively measure how the distributed implementation of minimax adaptive controller fair against the nominal distributed $H_{\infty}$ controller which assumes a nominal value of $a_{i}$ for all nodes $i \in \mathcal{V}$.

\subsection{Results \& Discussion}

\begin{figure}
\centering
\includegraphics[scale=0.135]{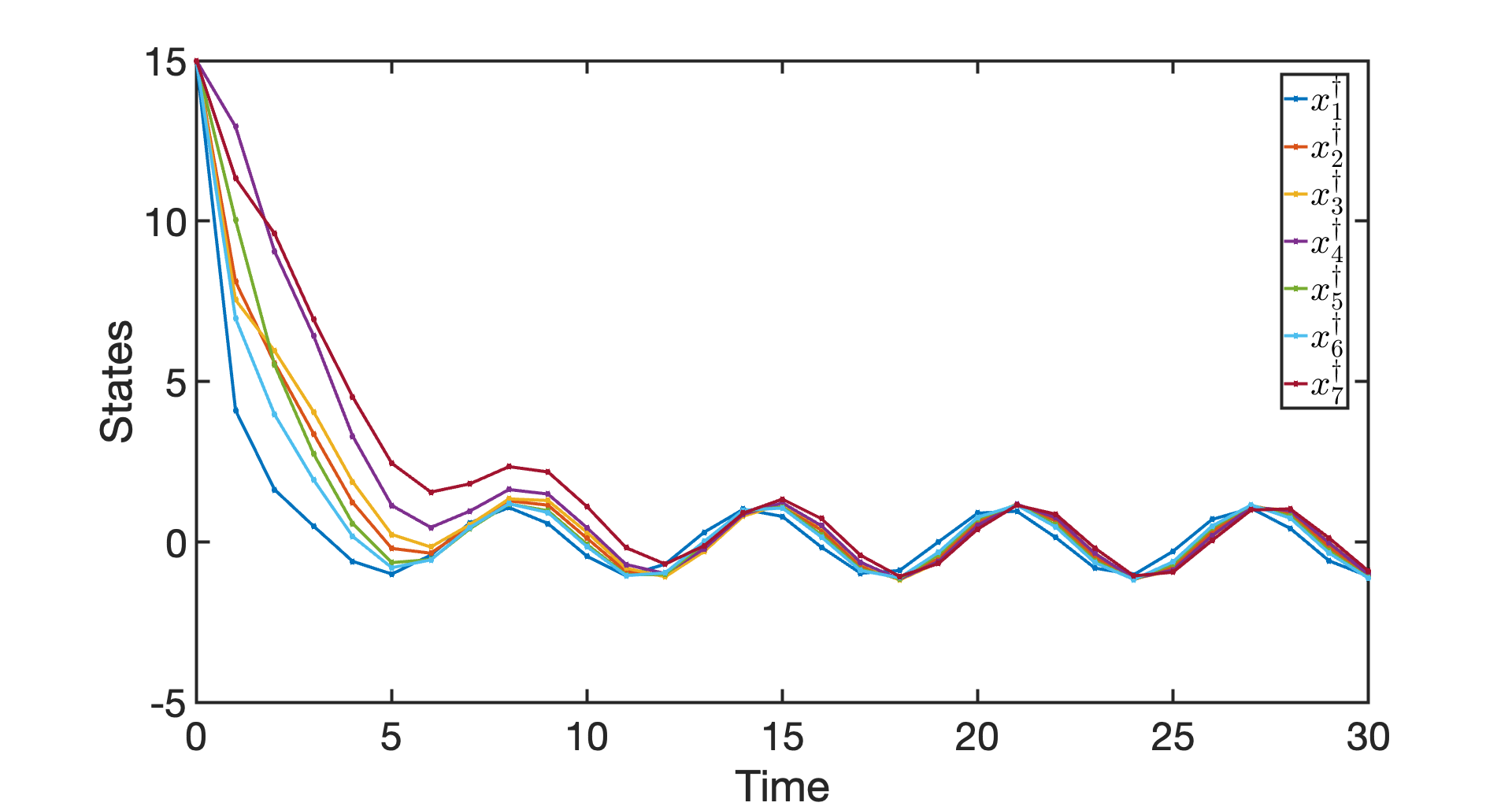}
\caption{The regulated state trajectories from the distributed minimax adaptive controller with sinusoidal disturbances are shown here.}
\label{fig_states_minmax}
\end{figure}

\begin{figure}
\centering
\includegraphics[scale=0.135]{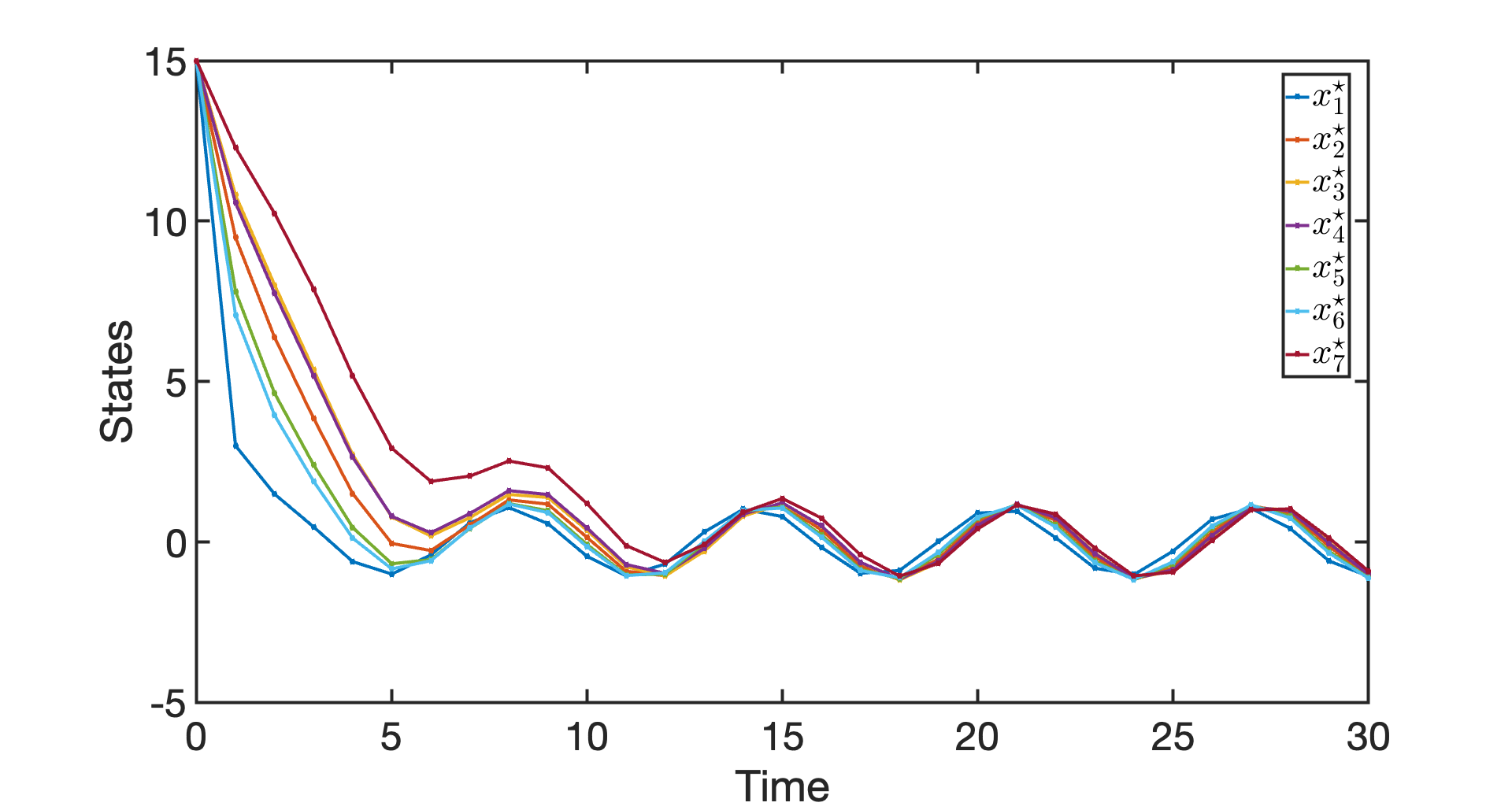}
\caption{The regulated state trajectories from the distributed $H_{\infty}$ controller with sinusoidal disturbances are shown here.}
\label{fig_states_h_infty}
\end{figure}

\begin{figure}
\centering
\includegraphics[scale=0.135]{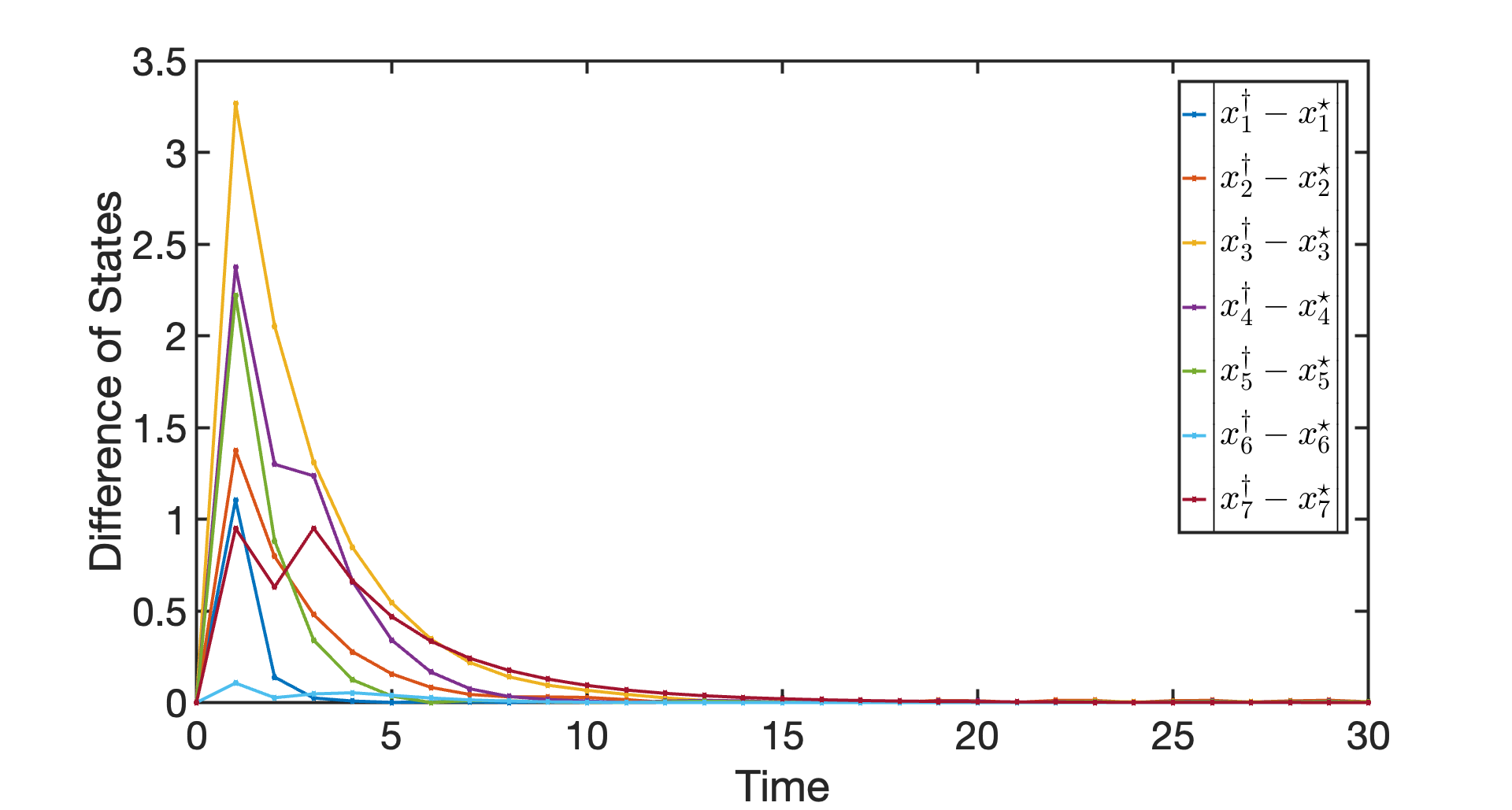}
\caption{The absolute difference between the state trajectories from the distributed minimax adaptive controller and the distributed $H_{\infty}$ controller with sinusoidal disturbances are shown here.}
\label{fig_states_difference}
\end{figure}

\begin{figure}
\centering
\includegraphics[scale=0.135]{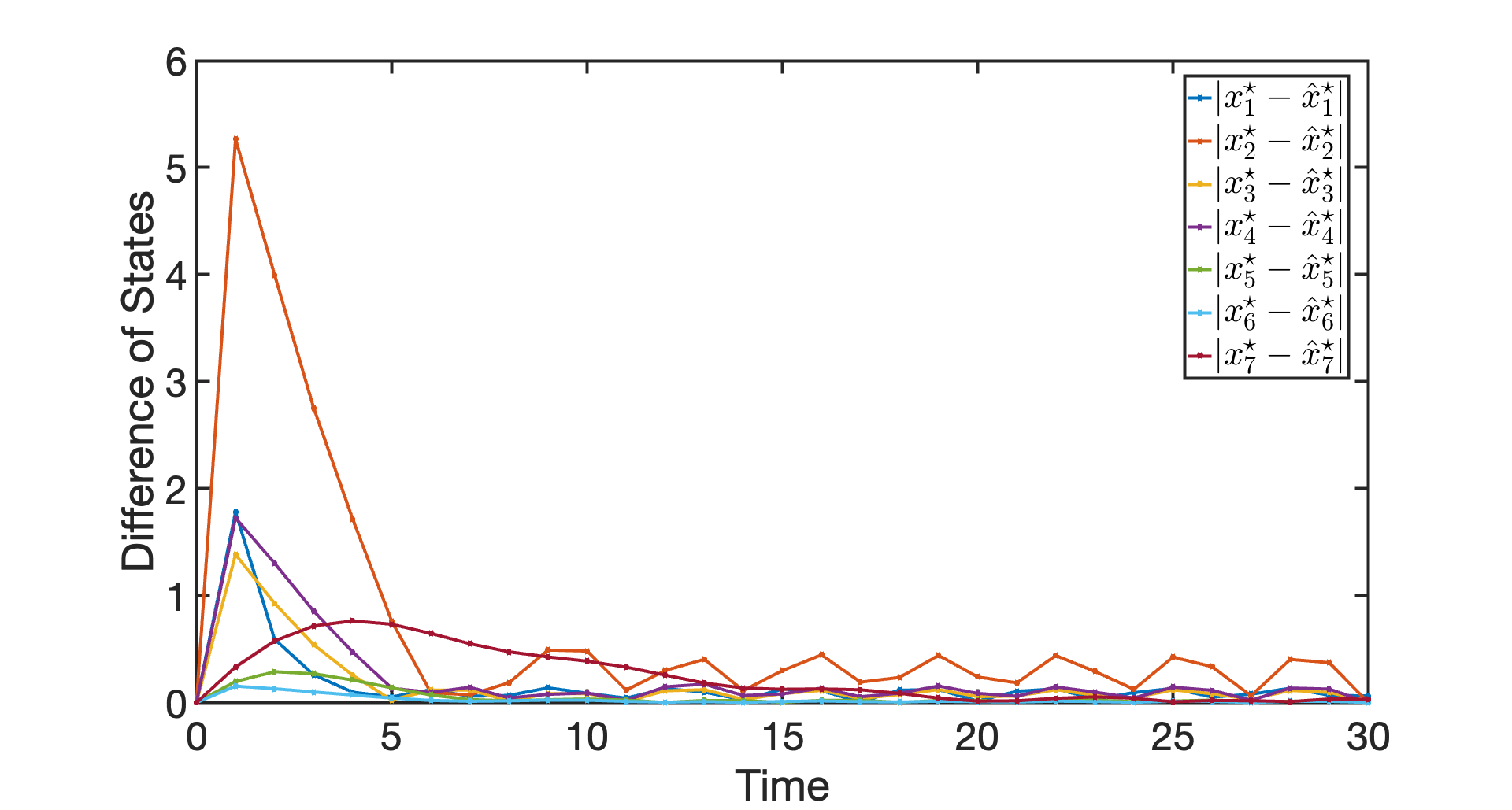}
\caption{The absolute difference between the state trajectories from the distributed $H_{\infty}$ controller that knows the true $a_{i} \in \mathbf{A_{i}}, \forall i \in \mathcal{V}$ and the nominal distributed $H_{\infty}$ controller which $\forall i \in \mathcal{V}$ assumes an $a_{i}$ value satisfying \eqref{eqn_communication_condition} to be its constant nominal value $\forall t \in \mathbb{N}$ under the sinusoidal disturbances are shown here. }
\label{fig_states_difference_nominal}
\end{figure}

It is evident from the Figure \ref{fig_states_minmax} that the distributed minimax adaptive controller was able to regulate the entire network states subject to the sinusoidal disturbance just like the distributed $H_{\infty}$ controller did by knowing apriori the true dynamics of each node in the network as shown in Figure \ref{fig_states_h_infty}. This means that the distributed minimax adaptive controller had figured out the uncertain dynamics from the recorded historical data and it has started to behave similar to that of the distributed $H_{\infty}$ controller as shown in Figure \ref{fig_states_difference}. In a way, Figure \ref{fig_states_difference} also depicts the regret experienced by the distributed minimax adaptive controller while compared against the distributed $H_{\infty}$ controller.

When we compare the system trajectory obtained using the nominal distributed $H_{\infty}$ controller for the true system against the system trajectory from the the distributed $H_{\infty}$ controller which knows the true $a_{i}$ for all nodes $i \in \mathcal{V}$ as shown in Figure \ref{fig_states_difference_nominal}, we observe that our distributed implementation in in Figure \ref{fig_states_difference} does better than the nominal distributed $H_{\infty}$ controller which assumes a nominal value for $a_{i}$ for all $i \in \mathcal{V}$ for simplicity. This demonstrates that learning the model that best fits the disturbance trajectory in the least square sense does a better job than just assuming a nominal value for the local dynamics model.


\section{Conclusion}
\label{sec:conclusion}
A distributed implementation of minimax adaptive controllers for networked dynamical systems modeled by a finite set of linear models was presented. Based on the local information collected by each node in the network, the minimax adaptive controller was implemented by selecting the best model that minimized the disturbance trajectory. Our proposed distributed implementation scales linearly with the size of the network and facilitates easy implementation and behaves very similarly like the distributed $H_{\infty}$ controller once the system matrix is sufficiently estimated. 


\bibliographystyle{IEEEtran}
\bibliography{references}


\appendices


\end{document}